\documentclass[prl,aps,twocolumn]{revtex4-1}
\usepackage{graphicx} 
\usepackage[usenames]{color}
\usepackage{amsmath,amssymb}
\usepackage{gensymb}
\usepackage{natbib}
\usepackage{xcolor}
\def\strutdepth{\dp\strutbox}
\def\nw#1{\strut\vadjust{\kern-\strutdepth\vtop to0pt{\vss\hbox to\hsize
{\hskip\hsize\hskip5pt$\leftarrow$\hss\strut}}}{\em #1}}
\usepackage{rotating}

\begin{document}
\newcommand{\comE}[1]{\textcolor{blue}{#1}}
\newcommand{\comER}[1]{\textcolor{red}{#1}}

\title{Turning bacteria suspensions into a "superfluid"}
\author{H{\'e}ctor Mat{\'i}as L{\'o}pez $^{1}$, Jer{\'e}mie Gachelin$^{2}$, Carine Douarche $^{3}$, Harold Auradou$^{1}$, Eric Cl\'ement$^{2}$}
\affiliation{$^{1}$ Univ Paris-Sud, CNRS, F-91405. Lab FAST, B\^at 502, Campus Univ, Orsay, F-91405 (France).\\
 $^{2}$Physique et M{\'e}canique des Milieux H{\'e}t{\'e}rog{e}nes (UMR 7636 ESPCI /CNRS /Univ.~P.M.~Curie /Univ.~Paris-Diderot), 10 rue Vauquelin, 75005 Paris, France. \\
 $^{3}$Laboratoire de Physique des Solides, Universit{\'e} Paris-Sud, CNRS UMR 8502, F-91405 Orsay.}%

\begin{abstract}
The rheological response under simple shear of an active suspension of \textit{Escherichia coli} is determined in a large range of shear rates and concentrations. The effective viscosity and the time scales characterizing the bacterial organization under shear are obtained. In the dilute regime, we bring evidences for a low shear Newtonian plateau characterized by a shear viscosity decreasing with concentration. In the semi-dilute regime, for particularly active bacteria, the suspension display a "super-fluid" like transition where the viscous resistance to shear vanishes, thus showing that macroscopically, the activity of pusher swimmers organized by shear, is able to fully overcome the dissipative effects due to viscous loss.
%

\end{abstract}


\date{\today}
\maketitle
Owing to its relevance in medicine, ecology and its importance for technological applications, the hydrodynamics of active suspensions is at the centre of many recent fundamental studies \cite{Koch2011, Marchetti2013}. In nature, wide classes of living micro-organisms move autonomously in fluids at very low Reynold numbers \cite{Lauga2009}. Their motility stems from  a variety of propulsive flagellar systems powered by nanomotors. For bacteria such a \textit{bacillus subtilis} or \textit{E.coli}, the propulsion comes from the rotation of helix shaped flagella creating a propulsive force at the rear of the cell body \cite{Berg2004}. Consequently, many original fluid properties stem from the swimming activity \cite{Hatwalne2004,Toner2005, Wu2000, Sokolov2009, Leptos2009, Rafai2010, Mino2011}. Due to hydrodynamic interactions bacteria may produce mesoscopic patterns of collective motion sometimes called ``bio-turbulence''\cite{Dombrowski2004, Sokolov2007, Sokolov2012, Saintillan2012, Dunkel2013, Gachelin2014}. In a flow, these bacteria may organize spatially \cite{Rusconi2014} and under shear, for pusher-swimmers, the swimming activity yields the possibility to decrease the macroscopic viscosity to values below the suspending fluid viscosity \cite{Hatwalne2004}. In the dilute regime, kinetic theories via a simple account of the dominant long range hydrodynamic field \cite{Haines2009, Saintillan2010,Ryan2011} provide closed-forms for shear viscosity as a function of shear rate. Remarkably, at low shear rate, these theories predict a Newtonian plateau with a viscosity decreasing linearly with concentration \cite{Haines2009, Saintillan2010, Ryan2011}. On the other hand, phenomenological theories were also proposed to describe macroscopically active suspensions via a coupling of hydrodynamic equations with polar and/or nematic order parameters \cite{Hatwalne2004,Toner2005,Cates2008, Giomi2010, Furthauers2012, Marchetti2013}. A striking outcome of these theories is that for a set of coupling parameters rendering essentially a high swimming activity, a self-organized motive macroscopic flow may show up in response to shear \cite{Cates2008, Giomi2010,Furthauers2012}. This onset of a dissipation-less current is described in analogy with the super-fluidity transition \cite{Cates2008, Giomi2010} of liquids. 
Experimental evidences for viscosity reduction to values below the suspending fluid viscosity were brought for \textit{Bacillus subtilis} \cite{Sokolov2009} and \textit{E. coli} \cite{Gachelin2013} suspensions. However, no full rheological characterization (i.e. viscosity \textit{versus} shear rate) under steady and uniform shear exists. Moreover, these pioneering experiments did not provide evidence for the low shear viscous plateau which is at the core of all theoretical predictions in the dilute regime. Finally, the phenomenological predictions for the non-linear regime have remained so far unobserved. Noticeably, for unicellular algea, viewed as "puller" swimmers, the predicted low shear rate increase of viscosity was measured experimentally \cite{Rafai2010}.
In this letter, in addition to a full rheological characterization of an \textit{E. coli} suspension, we  provide, in the dilute  regime and at low shear rate, experimental evidences for a linear decrease of the apparent viscosity with bacteria concentration. We also explore regimes of higher concentration and describe the conditions where we observe a transition to a dissipation-free macroscopic flow.\\
%
\begin{figure}[h!]
\begin{center}
\includegraphics[scale=0.41]{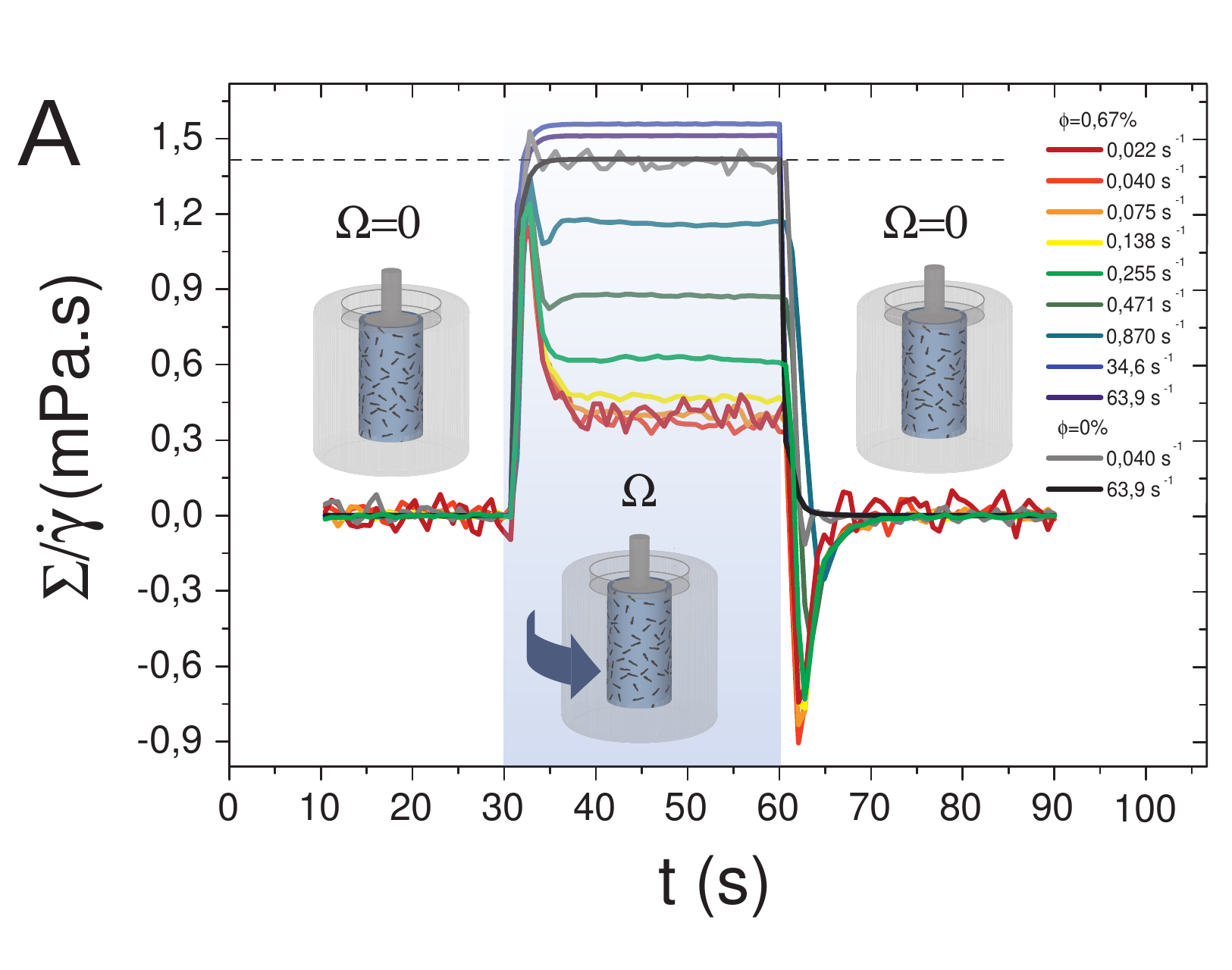}
\includegraphics[scale=0.425]{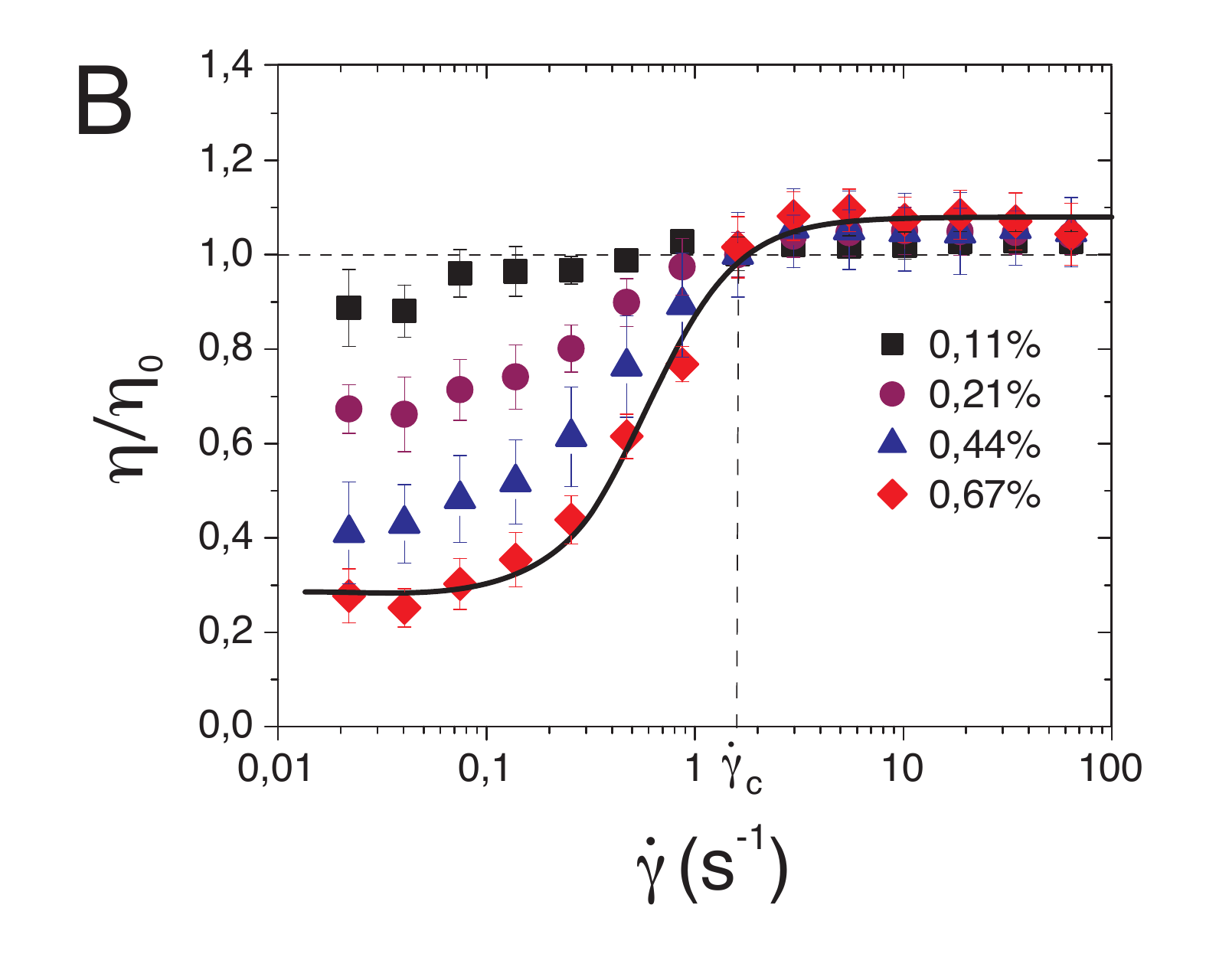}
\includegraphics[scale=0.425]{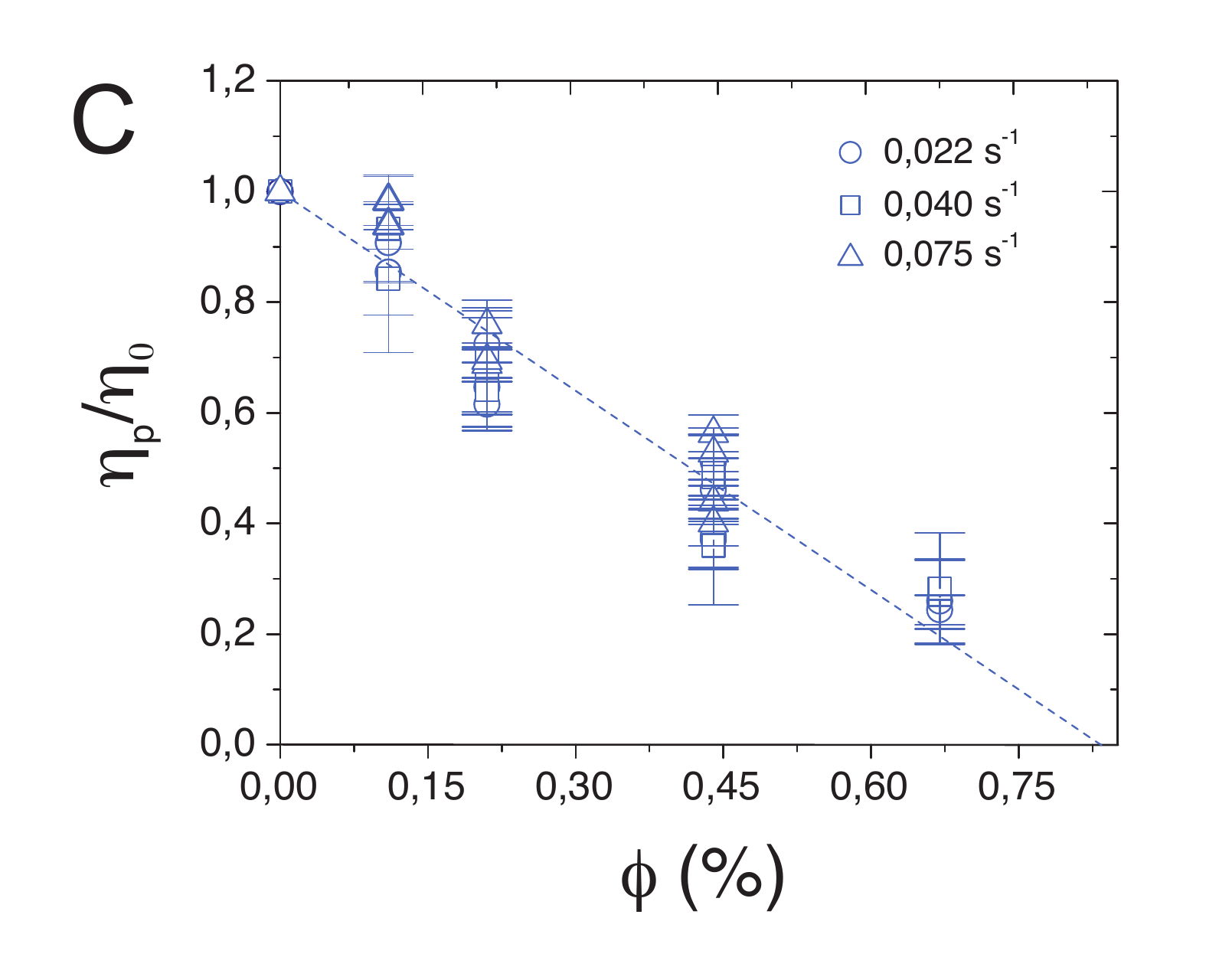}
\caption{
Shear stress response for an E.coli suspension (ATCC9637 strain, T=$25^\circ C$). (A) Shear stress $\Sigma$ rescaled by the applied shear rate $\dot{\gamma}$ during the rotation to display an effective viscosity $\Sigma/\dot{\gamma}$ in the sheared regime. Grey and black lines: fluid without bacteria ($\phi=0$). Colored solid lines: fluid with bacteria ($\phi = 0.67 \%$). Various $\dot{\gamma}$ are applied ranging from  64 s$^{-1}$ (dark blue line) down to 0.022 s$^{-1}$ (dark red line). (B) Relative viscosity $\eta/\eta_0$ averaged over three realisations as a function of $\dot{\gamma}$ ($\Box$: $\phi=0.11 \%$, $\bigcirc$: $\phi=0.21 \%$, $\bigtriangleup$: $\phi=0.44 \%$ and $\Diamond$: $\phi=0.67 \%$). The solid line is an adjustment by the Carreau law:  $\eta/\eta_0=1.08-0.795/(1+(\dot{\gamma}/0.6)^2)$. The vertical dashed line shows $\dot{\gamma}_c$, below this shear rate $\eta(\dot{\gamma})/\eta_0$ is less than $1$. (C) Values of the plateau viscosity $\eta_p/\eta_0$ as function of the bacteria volume fraction $\phi$ for very low shear rates ($\bigcirc$: $\dot{\gamma}$= 0.022 s$^{-1}$, $\Box$: $\dot{\gamma}$= 0.04 s$^{-1}$, $\bigtriangleup$: $\dot{\gamma}$= 0.075 s$^{-1}$).}
\label{fig:figure1}
\end{center}
\end{figure}


The active fluids considered here, are prepared out of two strains of wild type \textit{E. coli} (ATCC9637 and RP437) suspended into a minimal medium where the bacteria are still motile but do not divide. ATCC9637 is cultured overnight at 25 $^\circ$C in LB medium shaken at 240 rpm. RP437 is cultured overnight at 30 $^\circ$C and shaken at 240 rpm in M9 minimal medium supplemented with 1 mg/ml casamino acids and 4 mg/ml glucose. Next, the culture is washed twice by centrifugation ($2300$ g for $10$ min) and the cells are re-suspended into a motility medium containing $10$ mM potassium phosphate pH $7.0$, $0.1$ mM K-EDTA, $34$ mM K-acetate, $20$ mM sodium-lactate and $0.005\ \%$ polyvinylpyrrolidone (PVP-40). To avoid bacterial sedimentation, the suspension is mixed with Percoll (1 vol/1 vol). The bacteria concentration $n$ is represented by its volume fraction $\phi = n/V_b$, where $V_b$ is the bacteria body volume chosen as the classical value $ V_b=1 \mu m^3$.\\
Shear stress is measured in a low-shear couette rheometer (Contraves 30) designed especially for probing low-viscosity fluids. The inner bob (radius $R_i= 5.5$ mm, length 8 mm and underside cone angle 20$^\circ$ ) is suspended by a torsion wire into a cup (inner radius $R_i=6$ mm). The cup rotates at an angular rate $\Omega$ controlled by a computer. The corresponding shear rate is $\dot{\gamma}=\frac{\Omega R_0}{R_0-R_i}$. The central feature, making this instrument very precise for low stress measurements, is that the central bob is kept fixed by a feedback counter rotation of the suspending wire. The instrument measures the compensating torque required to keep the torsion wire at its null position. The torque is then converted into shear stress every $0.7$ s.\\ 
Importantly, due to the small surface area between the fluid and the air, the flux of $O_2$ is insufficient to compensate the amount of $O_2$ consumed by the bacterial activity.  
To avoid bacteria suffocation and consequently, a severe drop of activity, we supplement the suspension with L-Serine, an amino-acid allowing the bacteria to keep a significant swimming activity in absence of oxygen \cite{Adler1967, Douarche2009}. Therefore, in the early instants of the measurements, the bacteria are still in oxygenated conditions; but since they consume the oxygen, their mean velocity and diffusion coefficient decrease and stabilize within about $10$ min. Consequently, we observe a continuous increase in the suspension viscosity until a constant value is reached. Then, by metabolizing L-serine bacteria sustain a constant activity lasting for few hours (see Supporting Material (SM)). 
To obtain a full rheogram as displayed in Fig.\ref{fig:figure1}B, the following protocol is used. A volume ($1.25$ ml) of the suspension is poured into the rheometer's cup and then the bob is set into place. After $30~s$ of rest, the cup is rotated for $30~s$ at a steady state shear rate. The rotation is then stopped for $30$ s. These steps are repeated with increasing shear rate values. Once the highest $\dot{\gamma}$ is reached, the procedure is repeated in decreasing order to verify the reversibility of the viscous response. 

\begin{figure}[h!]
\begin{center}
\includegraphics[scale=0.425]{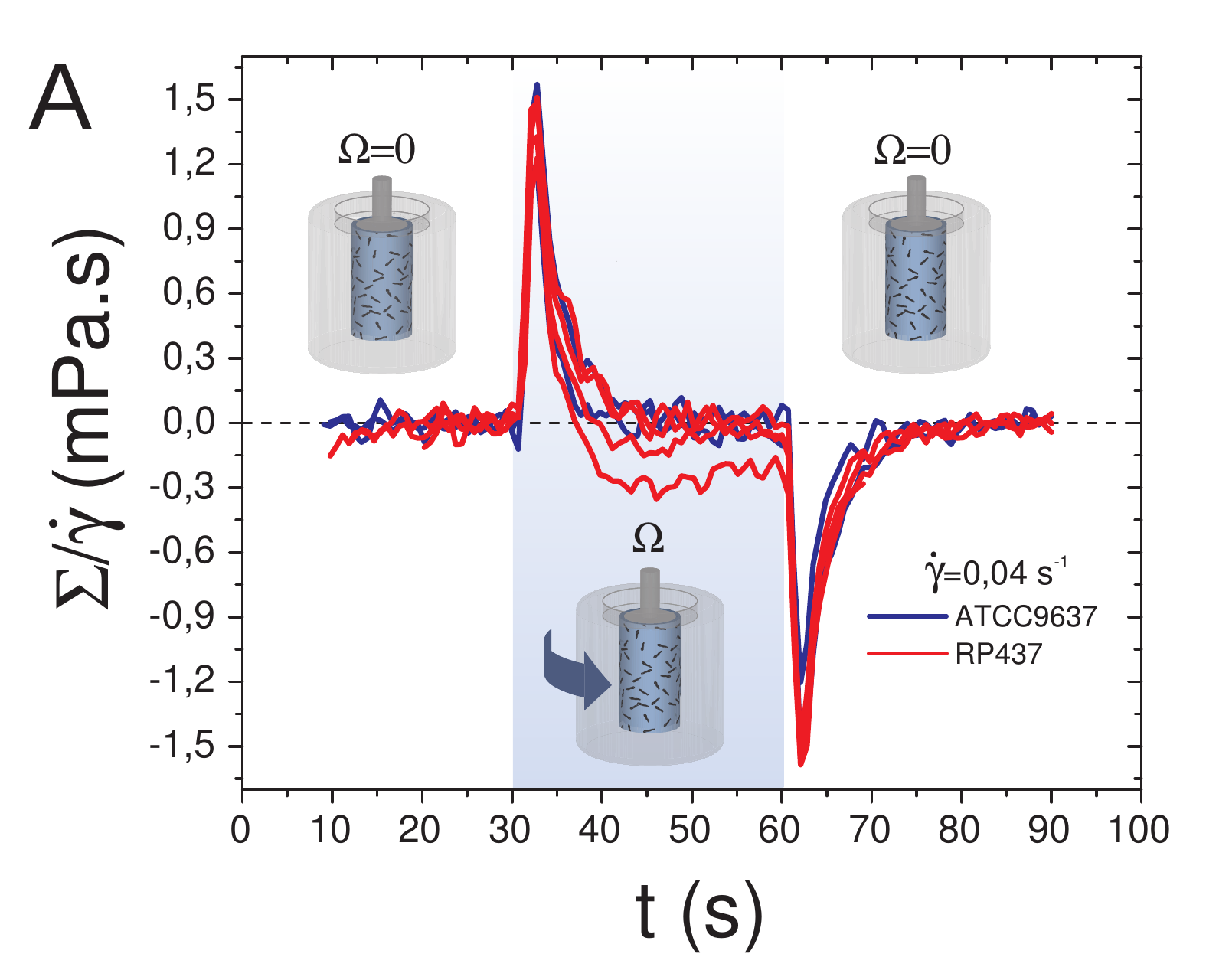}
\includegraphics[scale=0.425]{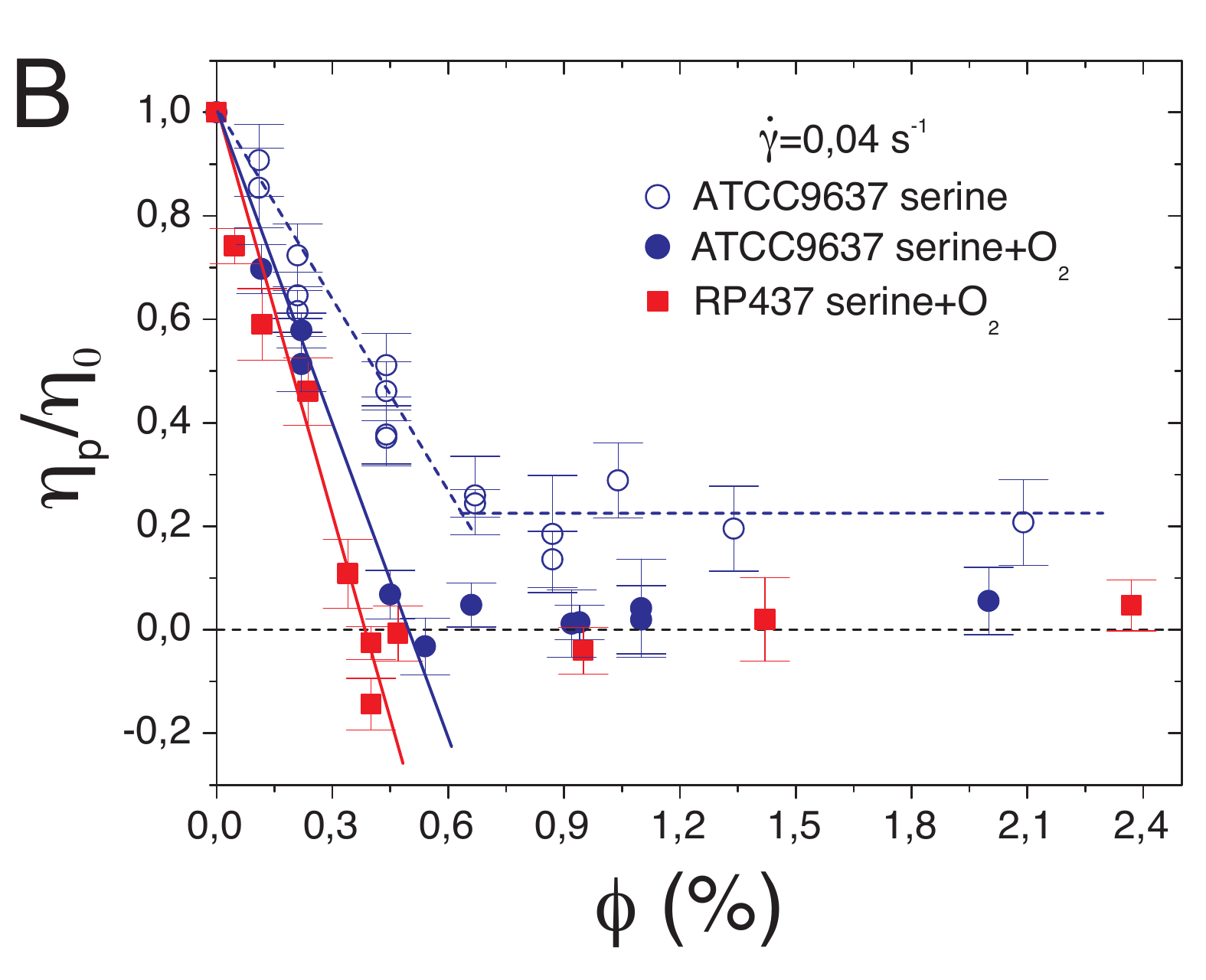}
\includegraphics[scale=0.425]{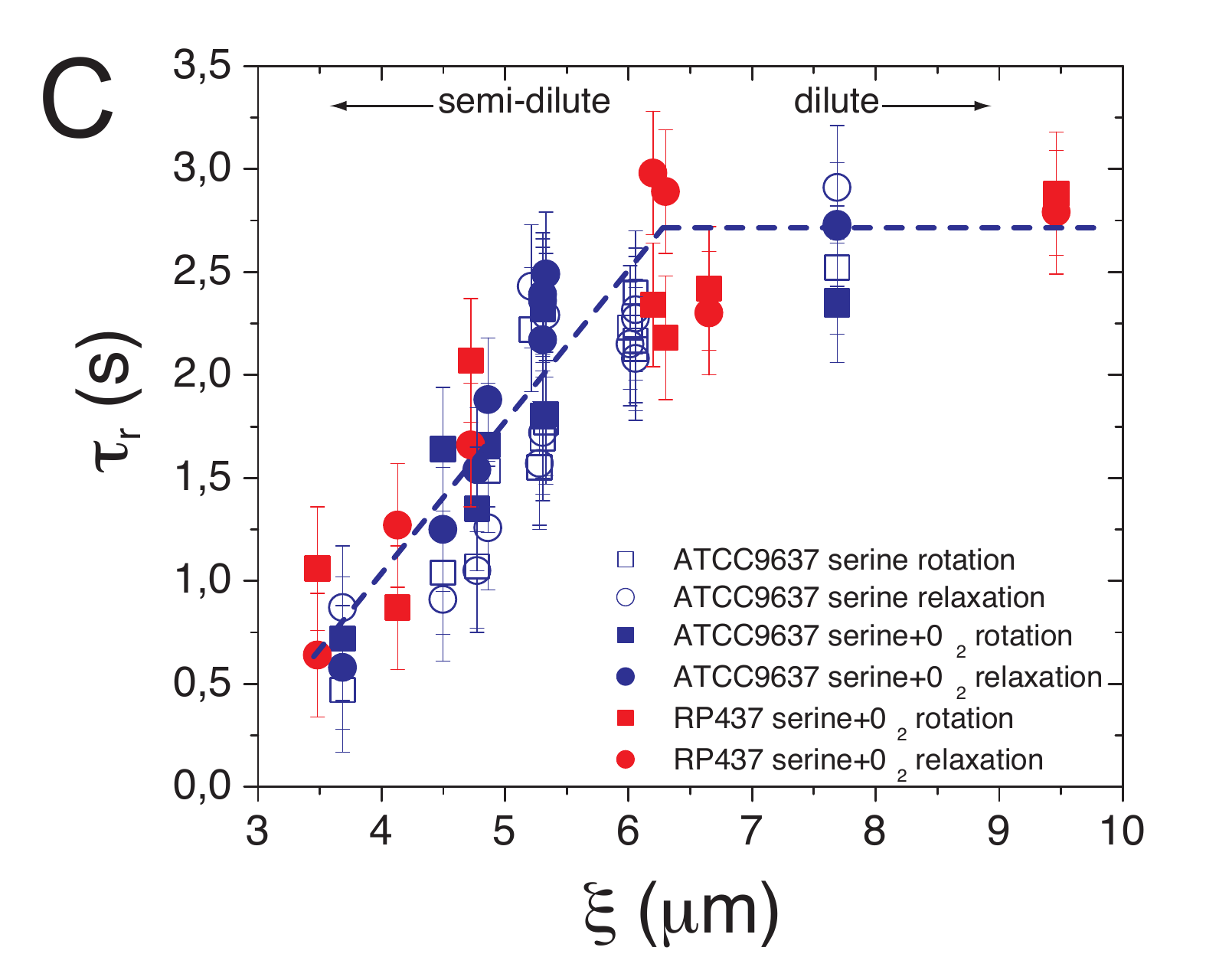}
\caption{(A) Shear stress response $\Sigma$ rescaled by the shear rate $\dot{\gamma}$  for the ATCC9637 strain (blue lines) and RP437 strain (red lines). All experiments are performed with $\dot{\gamma}$=0.04\ s$^{-1}$. In some cases, the stress reponse can reach negative values. (B) Variation of the viscosity $\frac{\eta_p}{\eta_0}$ as function of the volume fraction of bacteria $\phi$ in oxygenated conditions (filled symbols) and deoxygenated conditions (empty symbols). Dashed lines are meant as guides only. (C) Relaxation time $\tau_r$ obtained by adjusting exponentially the stress relaxation at the start of the shear (rotation) and at the end of the shear (relaxation) as function of the mean distance ($\xi=\phi^{-\frac{1}{3}}$) between bacteria (empty symbols: deoxygenated conditions, filled symbols: oxygenated conditions,  blue symbols: ATCC9637 strain, red symbols: RP437 strain).}
\label{fig:figure2}
\end{center}
\end{figure}

Rheology measurements were first performed to obtain viscosity at volume fractions between $\phi=0.1 \%$ and $0.67 \%$. 
In Fig. \ref{fig:figure1}A, we display the stress responses obtained for a suspension at a given $\phi$, for various shear rates. For the suspending fluid alone (a Newtonian fluid, $\eta_0=1.4 \textrm{mPa.s}$), stress-time responses, at the start or at the stop of the applied shear, are fast and correspond to the device compliance (grey and black lines). A similar behaviour is observed for the suspensions probed at high shear rates which moreover display a viscosity higher than $\eta_0$ as observed classically for suspensions of passive particles. However, at low shear rates, a strikingly different behaviour is observed. When shear starts, the stress jumps to the value measured in the absence of bacteria and then, after an exponential decrease, lasting for a few seconds, a steady effective viscosity - $\eta$ - is reached. When shear stops, the stress decreases abruptly and eventually changes sign. 
Finally, the stress relaxes exponentially to $0$ with a characteristic time $\tau_r^-$ not very different from the time scale, $\tau_r^+$, needed to reach a steady viscous response under shear (data on Fig.\ref{fig:figure2}C).
In this last stage, the bacterial motion induces a motive stress on the inner bob. 

Figure \ref{fig:figure1}B shows the suspension viscosity $\eta$ as a function of $\dot{\gamma}$ for different volume fractions ranging from  $\phi=0.11~\%$  ($1.1 \times 10^{9}$ bact/mL) up to $\phi= 0.67~\%$ ($6.7 \times 10^{9}$ bact/mL). 
We observe the three regimes predicted by the theories \cite{Haines2009,Saintillan2010}; 
(i) at high shear rates ($\dot{\gamma} >1 s^{-1}$),  the active contribution to viscosity is negligible and a Newtonian plateau appears akin to suspensions of passive particles of the same shape; (ii)  below a critical shear rate value $\dot{\gamma}_c \leq 1.5 \textrm{ s}^{-1}$, the suspension viscosity is lower than the suspending fluid viscosity; (iii) at low shear rates ($\dot{\gamma} \leq 0.1 \textrm{ s}^{-1}$), an "active viscous plateau" $\eta_p (\phi)$ appears. 
Futhermore, the theories also predict a linear dependance of $\eta_p$ with $\phi$ given by:
\begin{equation}
\frac{\eta_p}{\eta_0} = 1 +K(\frac{\tau}{t_c}) \phi 
\label{Eq1}
\end{equation}
where  $K \propto (A - B \frac{\tau}{t_c})$
; $t_c$ is the time taken by a bacterium to drag the fluid over its size and $\tau$ characterizes the directional persistence of a swimming trajectory \cite{Saragosti2012}; $A$ and $B$ depend solely on the bacterium shape \cite{Haines2009,Saintillan2010,Ryan2011}.
As shown in Fig.\ref{fig:figure1}C, $\eta_p(\phi)$ decreases linearly with the concentration as $\frac{\eta_p}{\eta_0}=1+K\phi $ with $K\simeq -120\pm 10$, as long as the shear rate is sufficiently low ({\it i.e.} in the range 0.022 - 0.075 s$^{-1}$). The experimental results are very consistent with active suspension theories for pusher swimmers in the dilute regime. 
Noticeably, within the framework of closed form theories established in the dilute regime, such a complete determination of the viscous response can be used to give an estimation of the microscopic bacterial activity (See SM for an explicit derivation of the dipolar strength and other microcopic parameters via the Saintillan \cite{Saintillan2010} kinetic model). 
%

We next compare the viscous response of the bacteria for two different activity levels. When both $O_2$ and L-serine are present in the suspension, one obtains an ''hyper-activated'' regime characterized by high values of bacteria diffusivity and swimming velocity ($D=7~10^{-11} m^2/s$ and $V_0 = 28 \mu m/s$ for ATCC9637). After $10$ min, when $O_2$ is consumed, the motility is maintained by the metabolizaton of L-serine. In this case one has a lower activity with $D=1.2~10^{-11} m^2/s$ and $V_0 = 20 \mu m/s$  (See SM Fig.2). 
The first regime, only lasts few minutes and the measurement were thus restricted to a single $\dot{\gamma}$ sufficiently low (here $0.04 s^{-1}$) to estimate $\eta_p$, the active plateau viscosity. 
In the "hyper-active" domain, $\eta_p$ is again found for $\phi<0.6 \%$ to decrease linearly with $\phi$ but with a larger slope $K' \simeq -200 \pm 3$ (see Fig. \ref{fig:figure2}B). 
This results demonstrates that the slope $K$ is strongly related to the bacterial activity. 
Next, these experiments were repeated with the second E-coli strain (RP437). 
A linear $\eta_p$ vs $\phi$ relation is also found but with a larger negative slope $K'' \sim -259 \pm 13$ (red squares in Fig. \ref{fig:figure2}B) demonstrating a correlation between the bacterial characteristics and the rheological response at low shear rate.\\
We finally increase the number of bacteria in the solution to values above $0.6 \%$. 
For the ATCC9637 strain in a medium that does not contain oxygen (empty symbols in Fig. \ref{fig:figure2}B), we observe that the viscosity becomes constant and independant of $\phi$  for $\phi>0.7 \%$ with $\eta_p/\eta_0 \sim 0.2$. In oxygenated conditions and for highly motile batches, we measure that the viscosity $\eta_p/\eta_0$ also reaches a constant value beyond $\phi \sim 0.5 \%$ and $\phi \sim 0.4 \%$ for ATCC9637 and RP437 respectively. For the three cases, the active plateau viscosity becomes thus independent of the concentration over a significant domain (from $\phi=0.6 \%$ up to $2.4 \%$). 
In addition under "hyper-activated" conditions, one observes a viscous response reaching zero (see Fig. \ref{fig:figure2}A), meaning that the local viscous dissipation is macroscopically entirely compensated by the swimming activity. Moreover, with the very active RP strain, negative values for $\eta_p$ (see Fig. \ref{fig:figure2}A and \ref{fig:figure2}B) could be obtained at the edge of the transition. 
These results point towards the idea of an organization process triggered by the shear flow which may last even when the shear ceases. 
In Fig. \ref{fig:figure2}C, we plot the relaxation times $\tau_r^+$ and $\tau_r^-$ as function of the mean distance $\xi$($=\phi^{-1/3}$) between bacteria. The transition between the dilute and the semi dilute regimes occurs here for a distance $\xi \simeq 6 \mu m$. 
In fact, below this length, the relaxation time is found to decrease linearly with $\xi$ (while the viscosity $\eta_p$ remains constant). The result seems to be consistent with the phenomenological picture proposed for active polar swimmer and with the clustering interaction recently observed in absence of flow \cite{Gachelin2014}, and suggests that collective effect becomes important for distances shorter than $6 \mu m$.

In conclusion our experiments show that the bacterial activity has a measurable 
influence on shear viscosity. We brought direct experimental evidence, at low shear rates, for an active viscous plateau which value decreases linearly with concentration (for $\phi<0.3 \%$). This confirms a central prediction for active pusher suspensions in the linear kinetic regime. 
The most striking feature of the rheological response is indeed the emergence of a viscous-less ''super-fluidity'' regime  ($\eta \sim 0$ or even lower). Presently, there is no \textit{ab initio} micro-hydrodynamic calculation describing the impact of bacteria interactions and possibly, the influence of collective organization on the macroscopic rheology. 
Therefore, one has to rely on phenomenological arguments to identify the possibly relevant macroscopic hydrodynamic contributions \cite{Cates2008, Marchetti2013}. In this framework, several authors \cite{Cates2008, Giomi2010, Furthauers2012} investigated the responses to shear of active polar particles and remarkably, they predicted the possibility of a transition to a ''zero-viscosity'' regime when the activity is increased. 
This result can also be cast in the framework of recent experimental works pointing out the possible use of bacterial motion 
to drive mechanical devices \cite{Hiratsuka2006, Sokolov2010, DiLeonardo2010}. Actually this report goes in this direction as we show, at least in principle, that rotational macroscopic power could be extracted from the swimming activity as for a rotatory motor \cite{Furthauers2012}. Finally, such a strong viscosity reduction may be a crucial element when considering macroscopic transport and particle dispersion in porous systems or in capillary networks, a central question to many applications involving bacterial fluids.

\begin{acknowledgments}
We acknowledge J.-P. Hulin, A. Lindner, A. Rousselet and D. Salin for usefull discussions and comments.
Correspondence and requests for materials
should be addressed to H. Auradou~(email: harold.auradou@u-psud.fr).
This work was partially supported by the Labex PALM.
\end{acknowledgments}

\vspace{-4 mm}

\end{document}


\section*{Supplemental Material of the manuscipt: "Turning bacteria suspensions into a "superfluid""}

\subsection*{Variation of the viscosity $\eta$ from oxygenated to deoxygenated conditions}

\begin{figure}[h!]
\centering
\includegraphics[width=3.5in]{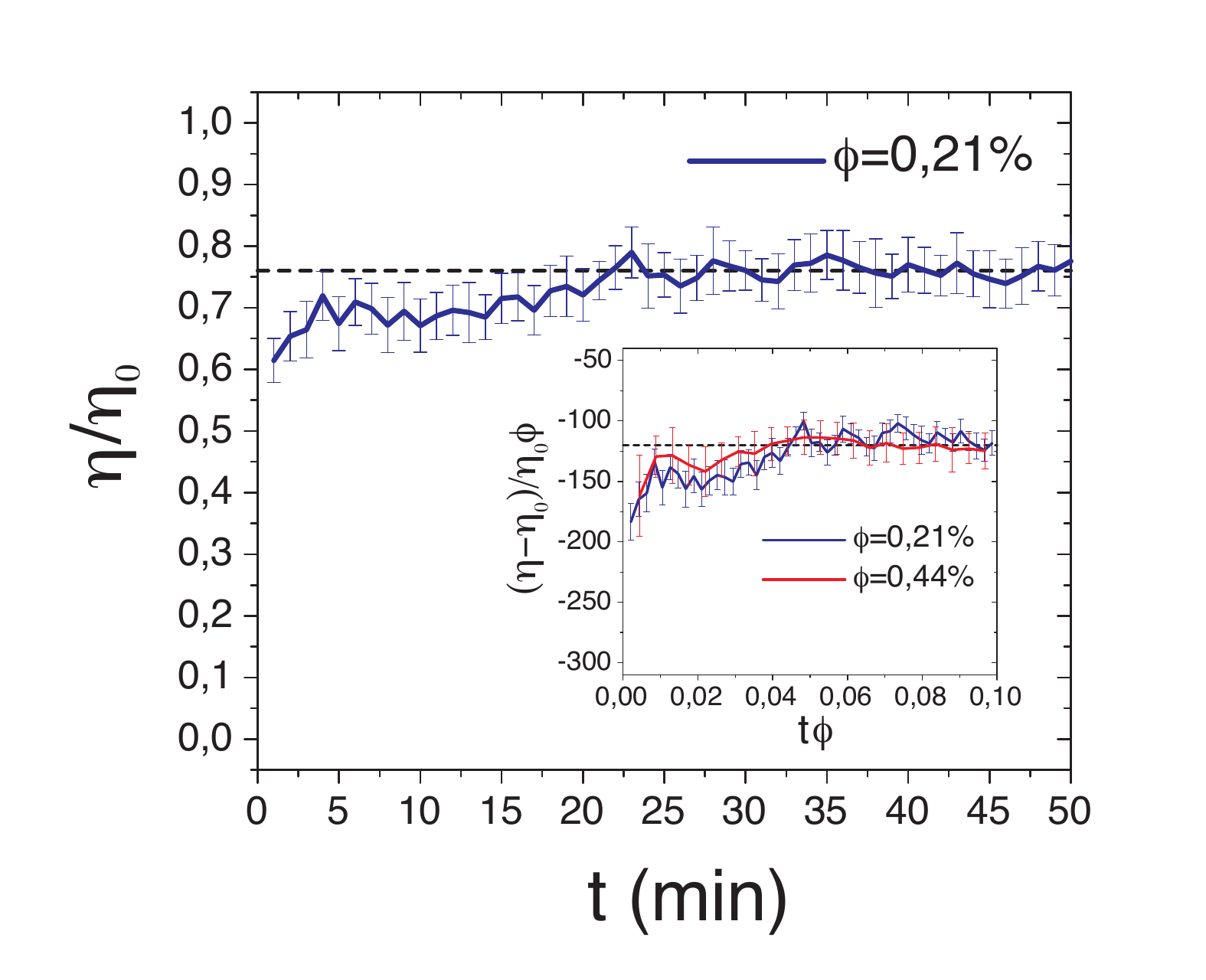}
\caption{{\bf{Variation of the viscosity $\eta$ from oxygenated to deoxygenated conditions at low shear $\dot{\gamma}=0.04\ s^{-1}$}}. The ratio of viscosities $\frac{\eta}{\eta_0}$ of a suspension containing a volume fraction $\phi=0.21\ \%$ of bacteria (ATCC9637) is initially of the order of $\sim$ 0.6 and reaches a constant value of $\sim$ 0.76 after 20 min. This increase is due to the consumption by the bacteria of the O$_2$ initially dissolved in the solution. After 20 min, the remaining activity is due to the metabolization of L-serine \cite{Douarche2009}. Inset: To demonstrate the relation between viscosity variation and O$_2$ consumption, the same measurement is performed with a solution containing twice the concentration of bacteria at $\phi=0.44\ \%$. The data collapse observed when the time $t$ is multiplied by the concentration $\phi$ demonstrates that it takes half of the time to reach the stabilized viscosity plateau suggesting a constant O$_2$ consumption rate. Thus, our data allow an estimation for the oxygen consumption rate by the bacteria. In this buffer  we find a rate of $\approx 2.5 \times 10^4$ dioxygen molecules/s/bacterium which is close to the value reported by Douarche and coworkers \cite{Douarche2009}.}
\label{fig:ED-figure1}
\end{figure}
\newpage
\subsection*{Evolution of the microscopic characteristics of bacteria motility}
\begin{figure}[h!]
\centering
\includegraphics[width=3.5in]{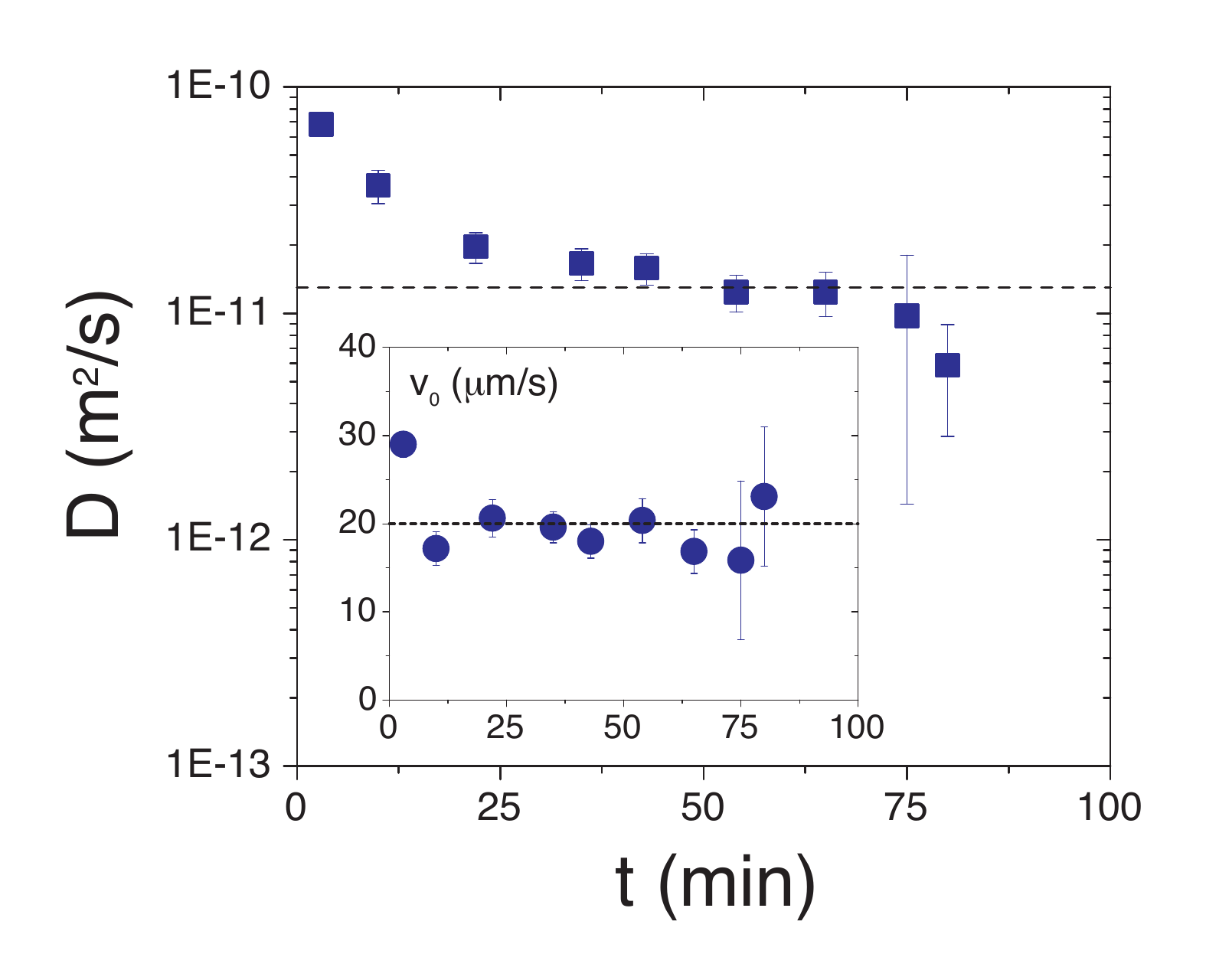}
\caption{{\bf{Evolution of the microscopic characteristics of bacteria motility.}} To confirm that the viscosity increase is due to a modification of the bacteria swimming characteristic, a suspension of bacteria ($\phi=0.3\ \%$, ATCC9637) is either placed in a sealed sample (\textit{i.e.} between two glass plates separated by a distance of $270 \mu$m and sealed by a GeneFrame gastight sealing system) or in a micro-fluidic channel made with an air permeable material. The bacteria are tracked at 125 fps and the mean square displacement is next computed as a function of time. Hence we extract the mean velocity $V_0(t)$ and the diffusion coefficient $D(t)$. In the micro-fluidic cell, where no shortage of O$_2$ occurs $D$ and $V_0$ are constant and close to $140\ \mu$m$^2$.s$^{-1}$ and $28\ \mu$m.s$^{-1}$ respectively. In the sealed sample, the bacteria consume the O$_2$ until exhaustion and turn deoxygented in presence of L-serine. This corresponds to a constant drop of both $D$ and $V_0$ which finally plateau at values : $D= 13\ \mu$m$^2$.s$^{-1}$ and $V_0=20\ \mu$m.s$^{-1}$.
The time characterizing the directional persistence $\tau_r=\frac{3D}{V_0^2}$ }
\label{fig:ED-figure2}
\end{figure}

\subsection*{Estimation of the dipolar strength $|\sigma_0|$, the thrust force $f$, and the effective size of a single bacterium $l$, via a slender body model.}
Here we  describe a method to obtain from the rheological responses -- the rheogram $\eta(\dot{\gamma} )$ and the stress relaxation time ($\tau_r$) -- the dipolar strength $|\sigma_0|$ of the bacteria. 
The latter can be seen as the work produced by one bacterium to move the fluid over its own size at a swimming velocity $V_0$. To do so, we use the theoretical and numerical results of Saintillan \cite{Saintillan2010} and we futhermore make the assumption that the relaxation time of the signal $\tau_r$ gives an estimation of the $\tau$ the characteristic time of directional persistence of the trajectories. In this article, Saintillan computes the contribution of active slender rod-like particles on the total viscosity $\eta$ for a suspending fluid of viscosity $\eta_0$. First, we use $\dot{\gamma_c}$ (\textit{i.e.} the shear rate for which $\frac{\eta(\dot{\gamma})}{\eta_0}$ starts to decrease below 1) to estimate $t_c$, the typical time for a bacteria to drag the fluid along its length $l$. \textbf{Fig. 4} in Saintillan's article \cite{Saintillan2010} displays for different values of $\frac{\tau}{t_c}$ the corresponding values of $\dot{\gamma_c} t_c$. In the actual range of operation of our experiment, we realize that a linear fit  of the numerical data  with $\frac{\tau}{t_c}\simeq 9\ \dot{\gamma_c} t_c$ allows to obtain a useful analytical relation between these variables. From the experimental measurements $\tau_r \sim 2.4 \pm 0.3$ s (See \textbf{Fig. 2c}) and $\dot{\gamma_c} \sim 1.5 \pm 0.25$ s$^{-1}$ (See \textbf{Fig. 1b}), we obtain a value $t_c=0.42 \pm 0.05$ s.\\

Finally, using the \textbf{Eq.(18)} of the article of Saintillan \cite{Saintillan2010} and considering the bacteria as slender rod-like particles (taking the Bretherton constant $\beta=1$),  we obtain the following analytical expression for $\eta_p$ :
\begin{equation}
\frac{\eta_p}{\eta_0}= 1+\frac{n |\sigma_0| t_c}{15 \eta_0}(4-\frac{\tau_r}{t_c})
\end{equation}
where $n=\frac{\phi}{v_b}$ is the number of bacteria per unit of volume and $v_b$ is the volume of a single bacteria. Since the empirical parameter $K$ is defined as $\frac{\eta_p}{\eta_0}=1+K \phi$, the dipolar strength $\sigma_0$ can be obtained from the microscopic parameters $\tau_r$, $t_c$  and $K$ : 
\begin{equation}
|\sigma_0|= \frac{15 K v_b \eta_0}{4 t_c - \tau_r}.
\end{equation}
Using the values $K \sim -120 \pm 10$, $\tau_r \sim 2.4 \pm 0.3 s$ and $t_c \sim0.42 \pm 0.05 s$ measured for the ATCC9637 strain when supplemented by L-Serine and $v_b=1\ \mu$m$^3$, we obtain : $|\sigma_0|= 3.8 \pm 1.0\ 10^{-18}$ Pa.m$^{3}$
By using the expression given by Saintillan \cite{Saintillan2010}: $t_c=\frac{\pi}{6}\eta_0\frac{l^3}{\sigma_0 ln(\frac{2l}{d})}$ where $l$ and $d$ are the dipolar length and the diameter of the bacterium respectively, we find that the corresponding dipolar length $l \sim 20.7 \pm 3\ \mu$m corresponding to the flagella bundle length plus the \textit{E. coli} body size \cite{Berg2004, Qian2007}. The dipole force $f=|\sigma_0|/l$ can also be estimated. For our measurements, we find a value of $f \sim 0.23 \pm 0.04$ pN. This agrees with recent experiments for swimming \textit{E. coli} reporting flagella thrust forces $f \sim 0.57$ pN \cite{Chattopadhyay2006}, $f \sim 0.41$ pN \cite{Darnton2007} and $f \sim 0.42$ pN using  PIV measurements \cite{Drescher2011}.\\

From the previous slender-body model, one can also predict (everything else being constant) the ratio of two slopes $K$ and $K'$ corresponding to two different swimming velocities $V_0$ and $V'_0$ when bacteria swim with and without oxygen respectively. Since 
$K = \frac{|\sigma_0|t_c}{15 \eta_0 v_b} (4- X)$, where $X=\frac{\tau_r}{t_c}$ and $t_c\propto 1/V$ then for the two bacteria velocities $V_0$ and $V'_0$, we have $\frac{K'}{K} = \frac{4-\frac{V'_0 X}{V_0}}{4- X}$.
For the ATCC9637 strain, under oxygenated conditions we have $V_0=20\ \mu$m.s$^{-1}$ and $X=\frac{\tau_r}{t_c} = 5.6 \pm 0.6$. When the cell activity is maintained by the metabolization of both O$_2$ and L-Serine (oxygenated condition), we measure a velocity $V'_0=28\ \mu$m.s$^{-1}$ and no significant change of $\tau_r$ is observed (See \textbf{Fig. 2c}). The ratio of the slopes is thus: $\frac{K}{K'} \simeq 0.41 \pm0.08$. This results is in agreement with our experimental measurement of the ratio: $\frac{K}{K'} = 0.56 \pm 0.05$.\\ 

Other models can be considered to estimate the microscopic property of the bacteria from rheological measurements. Haines and coworkers \cite{Haines2009} for instance considered the cell as a rigid prolate spheroid and Ryan {\it {et al.}} \cite{Ryan2011} as a point-like dipole. They both predict a Carreau-like variation of the viscosity of the form:
\begin{equation}
\frac{\eta}{\eta_0}= 1 - \phi \frac{\tau_r V_0}{l}\frac{1}{(1+(\frac{\dot{\gamma}}{\tau_r})^2)}
\end{equation} 
where $l$ is a typical length associated to the bacteria. Our experimental observations do agree with this type of variation (see solid line in \textbf{Fig. 1b}).

